\documentstyle[12pt]{article}

\catcode`\@=11
\long\def\@makefntext#1{
\protect\noindent \hbox to 3.2pt {\hskip-.9pt  
$^{{\ninerm\@thefnmark}}$\hfil}#1\hfill}		

\def\@makefnmark{\hbox to 0pt{$^{\@thefnmark}$\hss}}  
	
\def\ps@myheadings{\let\@mkboth\@gobbletwo
\def\@oddhead{\hbox{}
\rightmark\hfil\ninerm\thepage}   
\def\@oddfoot{}\def\@evenhead{\ninerm\thepage\hfil
\leftmark\hbox{}}\def\@evenfoot{}
\def\sectionmark##1{}\def\subsectionmark##1{}}

\renewcommand{\thefootnote}{\fnsymbol{footnote}}

\newcounter{sectionc}\newcounter{subsectionc}\newcounter{subsubsectionc}
\renewcommand{\section}[1] {\vspace*{0.6cm}\addtocounter{sectionc}{1} 
\setcounter{subsectionc}{0}\setcounter{subsubsectionc}{0}\noindent 
	{\normalsize\bf\thesectionc. #1}\par\vspace*{0.4cm}}
\renewcommand{\subsection}[1] {\vspace*{0.6cm}\addtocounter{subsectionc}{1} 
	\setcounter{subsubsectionc}{0}\noindent 
	{\normalsize\it\thesectionc.\thesubsectionc. #1}\par\vspace*{0.4cm}}
\renewcommand{\subsubsection}[1]
{\vspace*{0.6cm}\addtocounter{subsubsectionc}{1}
	\noindent {\normalsize\rm\thesectionc.\thesubsectionc.\thesubsubsectionc. 
	#1}\par\vspace*{0.4cm}}

\newcounter{appendixc}
\newcounter{subappendixc}[appendixc]
\newcounter{subsubappendixc}[subappendixc]

\renewcommand{\appendix}[1] {\vspace*{0.6cm}
        \refstepcounter{appendixc}
        \setcounter{figure}{0}
        \setcounter{table}{0}
        \setcounter{equation}{0}
        \renewcommand{\thefigure}{\Alph{appendixc}.\arabic{figure}}
        \renewcommand{\thetable}{\Alph{appendixc}.\arabic{table}}
        \renewcommand{\theappendixc}{\Alph{appendixc}}
        \renewcommand{\theequation}{\Alph{appendixc}.\arabic{equation}}
        \noindent{\bf Appendix \theappendixc #1}\par\vspace*{0.4cm}}

\def\abstracts#1{{
	\centering{\begin{minipage}{12.2truecm}\footnotesize\baselineskip=12pt\noindent
	\centerline{\footnotesize ABSTRACT}\vspace*{0.3cm}
	\parindent=0pt #1
	\end{minipage}}\par}} 

\renewenvironment{thebibliography}[1]
	{\begin{list}{\arabic{enumi}.}
	{\usecounter{enumi}\setlength{\parsep}{0pt}
\setlength{\leftmargin 1.25cm}{\rightmargin 0pt}
	 \setlength{\itemsep}{0pt} \settowidth
	{\labelwidth}{#1.}\sloppy}}{\end{list}}

\newcounter{itemlistc}
\newcounter{romanlistc}
\newcounter{alphlistc}
\newcounter{arabiclistc}

\newcommand{\fcaption}[1]{
        \refstepcounter{figure}
        \setbox\@tempboxa = \hbox{\footnotesize Fig.~\thefigure. #1}
        \ifdim \wd\@tempboxa > 6in
           {\begin{center}
        \parbox{6in}{\footnotesize\baselineskip=12pt Fig.~\thefigure. #1}
            \end{center}}
        \else
             {\begin{center}
             {\footnotesize Fig.~\thefigure. #1}
              \end{center}}
        \fi}

\newcommand{\tcaption}[1]{
        \refstepcounter{table}
        \setbox\@tempboxa = \hbox{\footnotesize Table~\thetable. #1}
        \ifdim \wd\@tempboxa > 6in
           {\begin{center}
        \parbox{6in}{\footnotesize\baselineskip=12pt Table~\thetable. #1}
            \end{center}}
        \else
             {\begin{center}
             {\footnotesize Table~\thetable. #1}
              \end{center}}
        \fi}

\def\@citex[#1]#2{\if@filesw\immediate\write\@auxout
	{\string\citation{#2}}\fi
\def\@citea{}\@cite{\@for\@citeb:=#2\do
	{\@citea\def\@citea{,}\@ifundefined
	{b@\@citeb}{{\bf ?}\@warning
	{Citation `\@citeb' on page \thepage \space undefined}}
	{\csname b@\@citeb\endcsname}}}{#1}}

\newif\if@cghi
\def\cite{\@cghitrue\@ifnextchar [{\@tempswatrue
	\@citex}{\@tempswafalse\@citex[]}}
\def\citelow{\@cghifalse\@ifnextchar [{\@tempswatrue
	\@citex}{\@tempswafalse\@citex[]}}
\def\@cite#1#2{{$\null^{#1}$\if@tempswa\typeout
	{IJCGA warning: optional citation argument 
	ignored: `#2'} \fi}}

 1
 1
 1

\font\ninerm=cmr9

\begin{document}

\begin{flushright}
\hfill{YUMS 95-24}\\
\hfill{SNUTP 95-118}\\
\hfill{KEK-TH-457}\\
\hfill{(November 1995)}
\end{flushright}
\vspace{0.5cm}

\centerline{\normalsize\bf FORM FACTORS OF SEMILEPTONIC $M_{l3}$ 
DECAYS\footnote{Based on Invited Talks given by C.S. Kim at International Symposium on Heavy Flavor and Electroweak Theory (16-19, August 1995, 
Beijing, China), and at Korean Theoretical Particle Physics Symposium 
(20-25, July 1995, Jeju Island, Korea). Both proceedings will be published by
World Scientific Publishing Company (1996).}}
\baselineskip=22pt

\centerline{\footnotesize C. S. Kim}
\baselineskip=13pt
\centerline{\footnotesize\it Department of Physics, Yonsei University}
\baselineskip=12pt
\centerline{\footnotesize\it Seoul 120-749, KOREA}
\centerline{\footnotesize E-mail: kim@cskim.yonsei.ac.kr}
\vspace*{0.3cm}
\centerline{\footnotesize and}
\vspace*{0.3cm}
\centerline{\footnotesize S. Y. Choi}
\baselineskip=13pt
\centerline{\footnotesize\it Theory Group, KEK, Tsukuba}
\baselineskip=12pt
\centerline{\footnotesize\it Ibaraki 305, JAPAN}
\centerline{\footnotesize E-mail: sychoi@theory.kek.jp}
\baselineskip=13pt

\vspace*{0.9cm}
\abstracts{We study in detail the kinematics of semileptonic
$M_{l3}$ ($M \rightarrow M^\prime l \nu$) decay,
considering only the new models
preserving all the gauge symmetries of the Standard Model (SM), 
and assuming that neutrinos are massless and left-handed as in the SM.
And we present a brief review of the recent theoretical progress on model independent constraints on the form factors of $M_{l3}$ decays 
using dispersion relations.
Finally we review a new parton model approach for semileptonic
$B \rightarrow D(D^*) l \nu$ decays,
by extending the inclusive parton model and by combining with 
the results of the heavy quark effective theory. 
We also obtain the slope of the Isgur-Wise function.}
 
\normalsize\baselineskip=15pt
\setcounter{footnote}{0}
\renewcommand{\thefootnote}{\alph{footnote}}
\section{Introduction}
In exclusive weak decay processes of hadrons, the effects of strong
interaction are encoded in hadronic form factors.
These decay form factors are Lorentz invariant
functions which depend on the momentum transfer $q^2$,
and their behaviors with varying $q^2$ are dominated by non-perturbative
effects of QCD.

Over the past few years, a great progress has been achieved 
in our understanding of the exclusive semileptonic decays 
of heavy flavors to heavy flavors\cite{NEU}.
In the limit where the mass of the heavy quark is taken to infinity, 
its strong interactions become independent of its mass and spin, 
and depend only on its velocity. 
This provides a new $SU(2N_f)$ spin--flavor symmetry, 
which is not manifest in the theory of QCD. However, 
this new symmetry has been made explicit 
in a framework of the heavy quark effective theory (HQET)\cite{HQ}. 
In practice, the HQET and this new symmetry relate 
all the hadronic matrix elements of 
$B \rightarrow D$ and $B \rightarrow D^*$ semileptonic decays, 
and all the form factors can be reduced to a single universal function, 
the so-called Isgur-Wise (IW) function\cite{HQ,IS},
which represents the common non-perturbative dynamics of 
weak decays of heavy mesons. 
However, the HQET cannot predict the values of the IW function 
over the whole $q^2$ range, though the normalization of the IW
function is precisely known in the zero recoil limit.
Hence the extrapolation of $q^2$ dependences of the IW function 
and of all form factors is still model dependent and the source of 
uncertainties in any theoretical model. 
Therefore, it is strongly recommended to determine
hadronic form factors of $M_{l3}$ ($M \rightarrow M^\prime l \nu$) 
decay more reliably, when we think
of their importance in theoretical and experimental analyses.

Although the Standard Model (SM) has been an enormous success in explaining
all the known experimental data, it is generally anticipated that there is
new physics in higher energy regions. One of the long-shot efforts
to exploring such new physics could be searches for CP violation
(or T violation) outside of
the Cabibbo-Kobayashi-Maskawa (CKM) paradigm\cite{ckm}.
It is well known\cite{Sakurai} that measuring a component of muon 
polarization normal to the decay plane in $K_{\mu 3}$
($K^+ \rightarrow \pi^0 \mu^+ \nu$) decay would signal T violation.
It is also expected that the CKM phase does not induce
the perpendicular muon polarization in $K_{\mu 3}$ decay.
Therefore, measurements\cite{Kuno} of these polarizations
could be clear signatures of physics beyond the SM.

In Section 2, we study in detail the kinematics of semileptonic $M_{l3}$
decay, considering only the new models
preserving all the gauge symmetries of the SM, and assuming that
neutrinos are massless and left-handed as in the SM.
And in Section 3, we present a brief review of the recent theoretical progress on model independent constraints on the form factors of $M_{l3}$ decays 
using {\it dispersion relations}.
Finally in Section 4, we review a new parton model approach for semileptonic
$B \rightarrow D(D^*) l \nu$ decays,
by extending the inclusive parton model and by combining with 
the results of the HQET. We also obtain the slope of IW function.

\section{Kinematics of semileptonic $M_{l3}$ decay}
The semileptonic hadron decays are very important in 
(i) determining the CKM matrix within the SM, 
and (ii) investigating new effects from physics
beyond the SM. In this report, considering only the new models
preserving all the gauge symmetries of the SM, and assuming that
neutrinos are massless and left-handed as in the SM, we concentrate 
on the semileptonic decay of a pseudo-scalar meson $M$ to a pseudo-scalar 
meson $M^\prime$ in the $M$ rest frame
\begin{eqnarray}
M(p)\rightarrow M^\prime(p^\prime)+l(k)+\nu(k^\prime),
\label{meson_decay}
\end{eqnarray}
where all the four-momenta are given in the parentheses.
In the SM the semileptonic decay process (\ref{meson_decay}) 
proceeds at quark level through a $W$ boson exchange, while 
the decay process may proceed through charged scalar/tensor 
exchanges in new physics beyond the SM as well. 

The decay amplitude for the process $M\rightarrow M^\prime l\nu$
can be parametrized in the factorized form as
\begin{eqnarray}
{\cal M}\propto G_FV\bigg[H^\mu L_\mu\bigg],
\label{decay_amplitude}
\end{eqnarray}
where $G_F$ is the Fermi decay constant and $V$ is a CKM
matrix element. While the leptonic current, $L^\mu$, is completely 
determined, the determination of the hadronic current,
$H^\mu$, is limited due to the complicated strong interactions. 
However, translation and Lorentz invariances force $H^\mu$ to be 
of the form for a scalar or vector exchange:
\begin{eqnarray}
H^\mu&=&(p+p^\prime)^\mu f_+(t)+(p-p^\prime)^\mu f_-(t),\nonumber\\
     &=&\bigg(p^\mu-\frac{p\cdot q}{q^2}q^\mu\bigg) f(t)
            +\frac{q^\mu}{q^2}d(t),
\label{parametrization}
\end{eqnarray}
where $q=p-p^\prime=k+k^\prime$ and $t=q^2$.
In Eq.~(\ref{parametrization}) we have introduced two different 
parametrizations, which are related as follows
\begin{eqnarray}
f(t)=2f_+(t),\qquad 
d(t)=(m^2-m^{\prime 2})f_+(t)+tf_-(t),
\end{eqnarray}
where $m(m^\prime)$ is the mass of the pseudo-scalar meson $M(M^\prime)$.
Clearly, the so-called scalar form factor $d(t)$ vanishes in the limit 
where the hadronic current is conserved. 
For our convenience, the latter form-factor set 
$(f(t), d(t))$ will be employed in the followings. 

It is now straightforward to calculate the absolute square of 
the decay amplitude (\ref{decay_amplitude}), which can be cast
into the form 
\begin{eqnarray}
|{\cal M}|^2\propto G^2_F|V_{12}|^2
    \bigg[T_f|f|^2+T_d|d|^2
         +T_{fd}{\cal R}(fd^*)+A_{fd}{\cal I}(fd^*)\bigg].
\end{eqnarray}
In general, the distributions, $T_i$ ($i=f,d,fd$) and $A_{fd}$ are 
functions of the four-momenta and the lepton polarization vector,
which can be explicitly determined.
First of all, we note that, when the summation over polarization
of the lepton is taken, the distribution, $A_{fd}$, vanishes and 
the other distributions are given by
\begin{eqnarray}
T_{f }&=&\frac{1}{8t^2}
     \bigg\{t(t-m^2_l)\left[(t+m^2-m^{\prime 2})^2-4m^2t\right]\nonumber\\
      & &\hskip 1cm -\left[(t+m^2_l)(t+m^2-m^{\prime 2})
                    -4tx\right]^2\bigg\},\nonumber\\
T_{d }&=&\frac{m^2_l}{2t^2}(t-m^2_l),\nonumber\\
T_{fd}&=&\frac{1}{4t^2}\bigg[(t+m^2_l)(t+m^2-m^{\prime 2})-4tx\bigg],
\end{eqnarray}
where $x=mE_l$ with the lepton energy $E_l$ in the $M$ rest frame. 
Secondly, we can see that three terms $|f|^2$, $|d|^2$, and 
${\cal R}(fd^*)$, are independently separable due to the different 
dependence of the three distributions on the variable $x$. 
Of course, in this case, the lepton mass $m_l$ should not be 
so small in order for the $|d|^2$ term to be measurable. 
On the other hand,  the differential decay width is given in terms 
of $t$ and $x$ by 
\begin{eqnarray}
{\rm d}\Gamma=\frac{1}{(2\pi)^3}\frac{|{\cal M}|^2}{16m^3}
              {\rm d}t{\rm d}x,
\end{eqnarray}
where $m^2_l\leq t\leq (m-m^\prime)^2$ and the scatter plot in
$t$ and $x$, which is called a Dalitz plot, has uniform phase
space density. 

With the lepton polarization into account, we find that
the distribution $A_{fd}$ in the $M$ rest frame is nothing but 
the so-called triple vector correlation
\begin{eqnarray}
A_{fd}=\frac{2m_l m}{t}
       \bigg[\vec{s}_l\cdot (\vec{k}\times \vec{p^\prime})\bigg],
\label{triple_product}
\end{eqnarray}
where $\vec{s}_l$ is the lepton polarization vector.

T-invariance requires that the relative phase of $f$ and $d$ 
should be the same or, in other words, ${\cal I}(fd^*)$ should be
zero. Conversely, any non-zero value of ${\cal I}(fd^*)$
would signal violation of T-violation\cite{Kuno}. 
It should be noted that {\it the T-violation effect can not be 
measured without the triple vector correlation} 
(\ref{triple_product}). Moreover, the T-violation effect is 
proportional to the mass of the lepton and requires the measurement 
of the the lepton polarization. 
While the polarization of the electron is almost not measurable, 
the polarization of the muon or tau leptons can be easily determined 
via the angular or energy distributions of their decay products. 
In light of these features, it is clear that in the $K$ meson case  
$K_{\mu 3}$ is by far more sensitive than $K_{e3}$, and in the $B$ 
meson case the $\tau$ lepton mode is most sensitive to ${\cal I}(fd^*)$. 
	
\input epsf.sty

\section{Model-independent constraints on the form factors}
The form factors at specific kinematic points can be approximately 
determined by the heavy quark symmetries for heavy meson decays 
and by the chiral symmetries for light meson decays. 
However, the event rate vanishes at the kinematic point so that 
the extrapolation of the form factor to other kinematical values 
QCDis required to determine the CKM matrix with good precision. 
Moreover, the form factor extrapolation is 
unavoidably accompanied by an uncertainty due to the choice of 
parametrization. Estimates of this uncertainty obtained by varying 
parametrizations suffer the same ambiguity as well. 
Recently, the possibility of obtaining model-independent bounds on 
the form factors $f(t)$ and $d(t)$ using dispersion relations
has revived much interest\cite{BGL,DT}. In this section
we present a brief review of the model-independent constraints 
on meson form factors using dispersion relations. 

The point\cite{BMD,Bo,Ok} is that the form factors $f(t)$ and $d(t)$ of  
the semileptonic decay process $M\rightarrow M^\prime l\nu_l$ 
can be constrained by the knowledge of the two-point function
\begin{eqnarray}
i\int {\rm d}^4x e^{iq\cdot x}
 \langle 0|{\rm T}(V^\mu(x)V^{\nu \dagger}(0)|0\rangle
  =-(g^{\mu\nu}t-q^\mu q^\nu)\Pi_T(t)+g^{\mu\nu}\Pi_L(t),
\label{two_point}
\end{eqnarray}
in the deep Euclidean region, where $V^\mu(x)$ denotes the vector
current responsible for the transition $M\rightarrow M^\prime$.
In order to render both sides of the relation (\ref{two_point}) 
finite we employ the once-subtracted dispersion relations
\begin{eqnarray}
\chi_{T,L}(Q^2)=\frac{\partial\Pi_{T,L}}{\partial q^2}|_{q^2=-Q^2}
               =\frac{1}{\pi}\int^\infty_0 {\rm d}t
                \frac{{\rm Im}\Pi_{T,L}(t)}{(t+Q^2)^2},
\label{once_subtract}
\end{eqnarray}
where the spectral functions ${\rm Im}\Pi_{T,L}(t)$ are defined by 
the relation
\begin{eqnarray}
&& -(g^{\mu\nu}t-q^\mu q^\nu){\rm Im}\Pi_T(t)
   +g^{\mu\nu}{\rm Im}\Pi_L(t)\nonumber\\
&&\mbox{ }\hskip 3cm =\frac{1}{2}\int\sum_\Gamma {\rm d}
   \rho_\Gamma (2\pi)^4 \delta(q-p_\Gamma)
   \langle 0|V^\mu(0)|\Gamma\rangle 
   \langle\Gamma| V^{\nu \dagger}(0)|0\rangle,
\label{relation}
\end{eqnarray}
with the summation over all possible hadron states $\Gamma$
of equal flavour quantum numbers and with an integral over
the phase space of allowed intermediate states. 
The hadron states $\Gamma$ might be resonances or open $MM^\prime$
states in a specific isospin channel determined by the quantum numbers 
of the relevant vector current $V^\mu$. In some cases the open $MM^\prime$
state is precisely the lowest hadronic state contributing to the absorptive
amplitude. Up to an overall Clebsh-Gordon coefficient $\eta$, the
function $\langle 0|V^\mu(0)|MM^\prime\rangle $ is {\it the same analytic
function} as $\langle M^\prime|V^\mu(0)|M\rangle $ which appears in
the decay $M\rightarrow M^\prime l\nu_l$, i.e.,
\begin{eqnarray}
\langle 0|V^\mu(0)|M\bar{M^\prime}\rangle 
 =\eta\Bigg\{\left(p^\mu-\frac{p\cdot q}{q^2}q^\mu\right) f(t)
             +\frac{q^\mu}{q^2}d(t)\Bigg\},
\end{eqnarray}
where the regime of the variable $t$ is $(m+m^\prime)^2\leq t\leq \infty$.

A judicious choice of the indices $\mu$ and $\nu$ makes the spectral 
functions a sum of positive definite terms, so one can obtain strict
inequalities by concentrating on the term with intermediate
states of $MM^\prime$ pairs. 
In particular, the intermediate open state $M\bar{M^\prime}$ 
gives the contribution 
\begin{eqnarray}
&&{\rm Im}\Pi_T(t)\geq \frac{\eta^2}{16\pi t}\sqrt{(t-t_-)(t-t_+)}
   \Bigg[\frac{|d(t)|^2}{t^2}
        +\left(1-\frac{t_-}{t}\right)\left(1-\frac{t_+}{t}\right)
         \frac{|f(t)|^2}{4}\Bigg],\nonumber\\
&&{\rm Im}\Pi_L(t)\geq \frac{\eta^2}{16\pi t^2}
          \sqrt{(t-t_-)(t-t_+)}|d(t)|^2, 
\label{inequality}
\end{eqnarray}
for $t\geq t_+$ with $t_+=(m+m^\prime)^2$ and $t_-=(m-m^\prime)^2$.
Clearly, the inequalities become tighter with more possible
contributions to the right side, which are always positive.
Eqs.~(\ref{once_subtract}) and (\ref{inequality})
allow us to obtain the inequalities on $\chi_{T,L}(Q^2)$, e.g.,
\begin{eqnarray}
\chi_L(Q^2)\geq \frac{\eta^2}{16\pi^2}\int^\infty_{t_+}{\rm d}t
   \frac{\sqrt{(t-t_-)(t-t_+)}}{t^2(t+Q^2)^2}|d(t)|^2\equiv J(Q^2).
\end{eqnarray}
The spectral function $\chi_{T,L}(Q^2)$ can be computed reliably 
from perturbative  for $Q^2$ far from the resonance region
and their lowest-order expressions from a quark one-loop diagram are 
\begin{eqnarray}
&&\chi_T(Q^2)=\frac{3}{4\pi^2}\int^1_0 {\rm d}x
  \frac{2x^2(1-x)^2}{x(1-x)Q^2+m^2_qx+m^2_{q^\prime}(1-x)},\nonumber\\
&&\chi_L(Q^2)=\frac{3}{4\pi^2}(m_q-m_{q^\prime})\int^1_0 {\rm d}x
  \frac{x(1-x)[m_qx-m_{q^\prime}(1-x)]}{x(1-x)Q^2+m^2_qx
               	+m^2_{q^\prime}(1-x)},
\label{one_loop}
\end{eqnarray}
where the quark $q$ ($q^\prime$) is inside $M$ ($M^\prime$). 

\epsfbox[85 280 590 539]{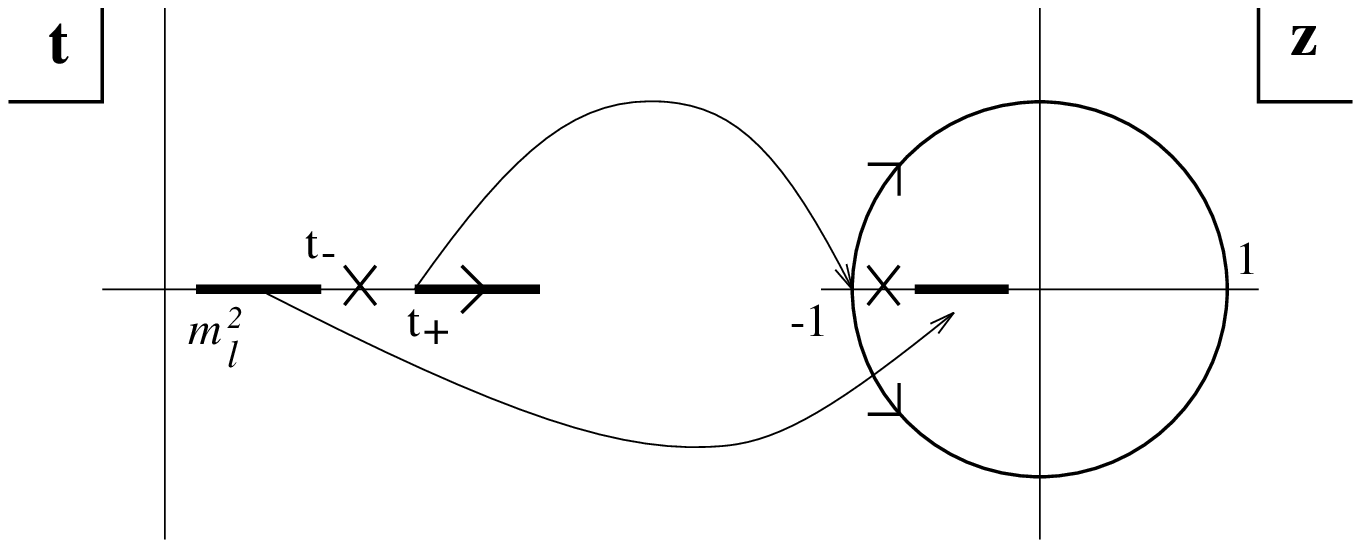}
\fcaption{Mapping of the cut $t$-plane into the unit disc by 
     the conformal transformation.}
\vskip 0.5cm

Using knowledge of the analytic structure of the form factors plus
the bounds (\ref{inequality}) one can derive model-independent
bounds on the form factors $f(t)$ and $d(t)$ in the physical region
of semileptonic decay, $m^2_l\leq t\leq (m-m^\prime)^2$. 
To this end let us map the complex $t$ plane onto the unit disk 
$|z|\leq 1$ by $\sqrt{(t_+-t)/(t_++Q^2)}=(1+z)/(1-z)$
(For the mapping, see Fig.~1), and then we can express $J(Q^2)$ 
as the norm squared of an analytic function, 
\begin{eqnarray}
J(Q^2)=||h||^2\equiv\frac{1}{2\pi}\int^{2\pi}_0{\rm d}\theta
       |\varphi({\rm e}^{i\theta})d({\rm e}^{i\theta})|^2,\qquad
       h(z)=\varphi(z)d(z),
\end{eqnarray}
on the unit disc of the complex $z$-plane with an explicitly determined 
function $\varphi(z)$.

With the inner product of functions defined on the unit disc as
\begin{eqnarray}
\langle g_1(z)|g_2(z)\rangle =\frac{1}{2\pi}\int^{2\pi}_0
    {\rm d}\theta g^*_1({\rm e}^{i\theta})g_2({\rm e}^{i\theta}),
\end{eqnarray}
we find that, for $g_1(z)=1/(1-\lambda^* z)$ and $g_2(z)=h(z)$,
$\langle g_1(z)|h(z)\rangle =d(\lambda)\varphi(\lambda)$,
$d$ can be determined\cite{Ok} with an appropriate 
$\lambda$ at any point of the physical region $m^2_l\leq t\leq t_-$ 
for the semileptonic decay $M\rightarrow M^\prime l\nu_l$. 
Furthermore, the $2\times 2$ positive semidefinite matrix of 
$A_{ij}=\langle g_i(z)|g_j(z)\rangle$, satisfying ${\rm det}(A_{ij})\geq 0$,
provides us with the model-independent bound on $d(\lambda)$
\begin{eqnarray}
|d(\lambda)|^2\leq \frac{J(Q^2)}{(1-|\lambda|^2)|\varphi(\lambda)|^2}
       \leq \frac{\chi_L(Q^2)}{(1-|\lambda|^2)|\varphi(\lambda)|^2}.
\end{eqnarray}
The bounds on the form factor $f(t)$ can be obtained through
the same approach as well.

To summarize, the  main aspects of the model-independent perturbative
QCD method are as follows.  
\begin{itemize}
\item The monemtum transfer squared $Q^2$ should be far from the 
      resonance region for reliable perturbative QCD calculations. 
      For example, for the heavy $B$ meson $Q^2=0$ can be 
      taken\cite{BGL} because of large resonance mass compared to the QCD 
      scale. However, for the light $K$ or $\pi$ mesons, the value 
      of $Q^2$ should be large\cite{BMD}. 
\item The spectral functions can be further saturated with more 
      contributions included. However, the inclusion of 
      resonances\cite{BGL,DT} can cause uncertainties due to no 
      precise information on their decay constants and masses.
      Of course, resonances with much larger than $t_+$ can be
      safely ignored. In this light, the $K\rightarrow\pi$ mode\cite{BGL} 
      is a very good system because of no resonances for $t\leq t_+$. 
      The $B\rightarrow \pi$ mode involves\cite{BGL} one resonance $B^*$, 
      while the $B\rightarrow D$ mode involves a lot of resonances, 
      which should be included for reliable predictions 
      on the form factors.
\item It is naturally expected that any theoretical or experimental 
      inputs\cite{BGL,DT} of the form factors $f(t)$ and $d(t)$ within 
      the physical region should give much stronger constraints on 
      the form factors. 
\end{itemize}

As briefly reviewed so far, the pertuarbatively calculable two-point 
functions in the deep Euclidean region provide us with model-independent 
bounds on the form factors of semileptonic meson decays and the constraints
can get much more stringent with more theoretical and experimental
determinations of the form factors at specific kinematic points. A
systematic investigation along this line is in progress\cite{CK}.

\section{Parton model approach for semileptonic $B \rightarrow D(D^*) l \nu$}
In this part we developed the parton model approach for exclusive 
semileptonic $B$ decays to $D,~D^*$, and predicted the $q^2$ dependences
of all form factors. For the details of this exclusive parton model approach, see Ref. Kim et al.\cite{KKL}.
Previously the parton model approach has been 
established to describe inclusive semileptonic $B$ decays\cite{Pas},
and found to give excellent agreements with experiments for electron
energy spectrum at all energies.

While many attempts\cite{WSB,KS} describing exclusive 
$B$ decays often take the pole-dominance ans$\ddot{\mbox {a}}$tze
as behaviors of form factors with varying $q^2$,
in our approach they are derived by the kinematical relations
between initial $b$ quark and final $c$ quark.
According to the Wirbel et al. model\cite{WSB}, 
which is one of the most popular
models to describe exclusive decays of $B$ mesons, 
the hadronic form factors are related to
the meson wavefunctions' overlap-integral in the infinite momentum
frame, but in our model they are determined by integral 
of the fragmentation functions, which are at least experimentally measurable.

We now develop the parton model approach for 
exclusive semileptonic decays of $B$ meson by 
extending the inclusive parton model, 
and by combining with the results of the HQET.
Theoretical formulation of this approach is, in a sense, closely related to
Drell-Yan process, while the parton model of inclusive $B$ decays is
motivated by deep inelastic scattering process.
And the bound state effects of exclusive $B$ decays are encoded into 
the hadronic distribution functions of partons inside an initial $B$ meson
and  of partons of a final state resonance hadron.
Then, the Lorentz invariant hadronic decay width can be obtained
using the hadronic distribution functions $f_{D,B}(x)$,
\begin{eqnarray}
E_B \cdot d\Gamma(B \rightarrow D(D^*)e\nu) = \int dx \int dy~
                      f_B(x)~E_b \cdot d\Gamma(b \rightarrow ce\nu)~f_D(y)~~.
\end{eqnarray}
The first integral represents the effects of motion of $b$ quark
within $B$ meson and the second integral those of $c$ quark within $D$ meson.
The variables $x$ and $y$ are fractions of momenta of partons 
to momenta of mesons,
\begin{eqnarray}
p_b = x p_B~,~~~~~~~~~ 
p_c = y p_D~,
\end{eqnarray}
in the infinite momentum frame.
The functions $f_B(x)$ and  $f_D(y)$ are the distribution function 
of $b$ quark inside $B$ meson, and the fragmentation function of $c$ quark 
to $D$ meson respectively.
Since the momentum fractions and the distribution functions are all defined 
in the infinite momentum frame, we have to consider the
colliderLorentz invariant quantity, ~$E \cdot d\Gamma$, to use at any other frame.

The distribution function can be identified with the fragmentation function
for a fast moving $b$-quark to hadronize into a $B$ meson in the infinite
momentum frame.
In general the distribution and fragmentation functions 
of a heavy quark ($Q=t,b,c$) in a heavy meson ($Q q$),
which are closely related by a time reversal transformation,
are of very similar functional forms, and peak both at large value of $x$.
Brodsky et al.\cite{Brod} have calculated 
the distribution function of a heavy quark,
which has the same form as Peterson's fragmentation function\cite{Pet}.
Therefore, here we follow the previous 
works\cite{Pas}
to use the Peterson's fragmentation function for both
distributions,  $f_B(x)$ and  $f_D(y)$. It has the functional form:
\begin{eqnarray}
f_Q(z) = N_Q z^{-1}~
     \left( 1-\frac{1}{z}-\frac{\epsilon_Q}{1-z} \right)^{-2}~~,
\end{eqnarray}
where $N_Q$ is a normalization constant, and $Q$ denotes $b$ or $c$ quark. 
This functional form is not purely ad hoc., but motivated
by general theoretical arguments,
that the transition amplitude for a fast moving heavy quark $Q$
to fragment into a heavy meson $(Qq)$ is proportional to 
the inverse of the energy transfer $\Delta E^{-1}$.
When we use this function, we have the advantage that 
the parameter $\epsilon_Q$ of this function has been determined 
from  high energy  experiments.
Thus we can replace uncertain wave functions of heavy mesons
by the experimentally measurable fragmentation functions.

{}From Lorentz invariance we write the matrix element of the decay 
$\bar{B} \rightarrow D e \bar{\nu}$ in the form
\begin{eqnarray}
<D|J_{\mu}|B> = f_+(q^2) (p_B+p_D)_{\mu} + f_-(q^2) (p_B-p_D)_{\mu}~~, 
\end{eqnarray}
and in terms of the HQET 
\begin{eqnarray}
<D(v')|J_{\mu}|B(v)> = \sqrt{m_B m_D}~(\xi_+(v \cdot v') (v+v')_{\mu} 
				     + \xi_-(v \cdot v') (v-v')_{\mu})~~.
\end{eqnarray}
Due to the conservation of leptonic currents, the form factors
multiplicated by $q_{\mu}$ do not contribute.
Then the hadronic tensor is given by
\begin{eqnarray}
H_{\mu \nu} &=& <D|J_{\mu}|B><D|J_{\nu}|B>^*
           \nonumber \\
            &=& 2~|f_+(q^2)|^2 ({p_B}_{\mu} {p_D}_{\nu} 
                              + {p_B}_{\nu} {p_D}_{\mu})~~,
\end{eqnarray}
and can be expressed by the IW function,
\begin{eqnarray}
H_{\mu \nu} = R^{-1} |\xi(v \cdot v')|^2  
             ({p_B}_{\mu} {p_D}_{\nu} + {p_B}_{\nu} {p_D}_{\mu})
                       \left( 1+{\cal O}(\frac{1}{m_Q}) \right)~~,
\end{eqnarray}
where 
\begin{eqnarray}
R=\frac{2\sqrt{m_B m_D}}{m_B+m_D}~~. \nonumber
\end{eqnarray}

By comparing the matrix element of the decay 
$\bar{B} \rightarrow D e \bar{\nu}$ in the parton model approach
and the matrix element in terms of the HQET,
we can define the function ${\cal F}(q^2)$ as
\begin{eqnarray}
{\cal F}(q^2) \equiv  \int dx~f_B(x) f_D(y(x,q^2)) ~xy^3(x,q^2)~~.
\label{fq2}
\end{eqnarray}
For given $q^2$ in our parton picture, the function ${\cal F}(q^2)$ 
measures the weighted transition amplitude, which is explicitly 
given by the overlap integral
of distribution functions of initial and final state hadrons.
Comparing the hadronic tensor of HQET and that of parton model approach, 
the IW function is calculated 
\begin{eqnarray}
|\xi(v \cdot v')|^2 \left( 1 + {\cal O}(\frac{1}{m_Q}) \right)
                 = 4 N \cdot R \cdot {\cal F}(v \cdot v')~~.
\end{eqnarray}
Finally the $q^2$ spectrum is given by
\begin{eqnarray}
\frac{d\Gamma(\bar{B} \rightarrow D e \bar{\nu})}{dq^2} 
  = \frac{{G_F}^2 |V_{cb}|^2}{96 \pi^3 m_B^3} 
{\cal F}(q^2) \left( (m_B^2-m_D^2+q^2)^2-4 m_B^2 q^2 \right)^{3/2}~~.
\end{eqnarray}
We can also obtain the decay spectrum for $B \rightarrow D^* l \nu$
\begin{eqnarray}
\frac{d\Gamma(\bar{B} \rightarrow D^* e \bar{\nu})}{dq^2} 
  &=& \frac{G_F^2 |V_{cb}|^2}{192 \pi^3 m_B^5} {\cal F}(q^2) 
      \left( (m_B^2-m_{D^*}^2+q^2)^2-4 m_B^2 q^2 \right)^{1/2}
      \nonumber \\
  & & ~~~
      \times \left[
        m_B^2 W_1(q^2)~((m_B^2-m_{D^*}^2+q^2)^2 -m_B^2 q^2) \right.
      \nonumber \\
  & & ~~~~~~~
      \left. + \frac{3}{2} m_B^2 W_2(q^2) ~(m_B^2-m_{D^*}^2+q^2) 
                 + 3 m_B^2 W_3(q^2) 
      \right]~~,
\end{eqnarray}
where
\begin{eqnarray}
W_1(q^2) &=& - N_1 \left( 1-\frac{2q^2}{(m_B+m_{D^*})^2} \right)~~,
        \nonumber \\
W_2(q^2) &=&  N_1 (m_B^2-m_{D^*}^2+q^2) 
               \left(1-\frac{2q^2}{(m_B+m_{D^*})^2}\right)
       - 2 N_3 q^2 \left( 1-\frac{q^2}{(m_B+m_{D^*})^2} \right)~~,
        \nonumber \\
W_3(q^2) &=& - N_1 m_B^2 q^2 \left( 1-\frac{2q^2}{(m_B+m_{D^*})^2} \right)
        \nonumber \\
    & &~~~~+ N_3 q^2 (m_B^2-m_{D^*}^2+q^2) 
         \left( 1-\frac{q^2}{(m_B+m_{D^*})^2} \right)
        \nonumber \\
    & &~~~~~~~~+ N_2 q^2 (m_B^2+m_{D^*}^2-q^2) 
         \left( 1-\frac{2q^2}{(m_B+m_{D^*})^2} \right)~~,
\end{eqnarray}
and ${\cal F}(q^2)$ is defined in (\ref{fq2}).

The result is plotted in Fig. 2, also
compared  with the recent CLEO data\cite{CLEO}.
The thick solid line is our model prediction 
with the parameters ($\epsilon_b =$0.004,  $\epsilon_c =$0.04)
for the heavy quark fragmentation functions\cite{Pet}, 
the thin solid line the Wirbel et al. model prediction\cite{WSB}, 
and the dotted line the K\"orner et al. model prediction\cite{KS}.
We also obtain the values of the slope parameter ${\hat \rho}$ within the 
parton model framework,
\begin{eqnarray}
 {\hat \rho}^2 = 0.582 - 0.896~~,
 \nonumber
\end{eqnarray}
which are compatible with the average value measured by experiments\cite{exp},
\begin{eqnarray}
 {\hat \rho}^2 = 0.87 \pm 0.12~~.
 \nonumber
\end{eqnarray} 
For more details of predictions from this exclusive parton model approach, 
see Ref. Kim et al.\cite{KKL}.

\setlength{\unitlength}{0.240900pt}
\ifx\plotpoint\undefined\newsavebox{\plotpoint}\fi
\sbox{\plotpoint}{\rule[-0.500pt]{1.000pt}{1.000pt}}%

\fcaption{$q^2$ spectrum in $B \rightarrow D^* e \nu$ decays.}

\section{Acknowledgements}
The work of CSK was supported 
in part by the Korean Science and Engineering  Foundation, 
Project No. 951-0207-008-2,
in part by Non-Directed-Research-Fund, Korea Research Foundation 1993, 
in part by the CTP, Seoul National University, 
in part by Yonsei University Faculty Research Grant 1995,  and
in part by the Basic Science Research Institute Program, Ministry 
of Education,  Project No. BSRI-95-2425. The work of SYC was
supported by the Japan Society for the Promotion of Science 
(No.~94024).

\end{document}